\documentclass{iau}

\usepackage{amsmath}
\usepackage{graphicx}
\usepackage{multirow}

\def\gtrsim{\lower 2pt \hbox{$\, \buildrel {\scriptstyle >}\over
{\scriptstyle \sim}\,$}}
\def\lesssim{\lower 2pt \hbox{$\, \buildrel {\scriptstyle <}\over
{\scriptstyle \sim}\,$}}

\def\xmm{{\sl XMM-Newton}}

\def\suzaku{{\sl Suzaku}}
\def\chandra{{\sl Chandra}}

\def\approxlt{\lower.2em\hbox{$\buildrel < \over \sim$}}
\def\approxgt{\lower.2em\hbox{$\buildrel > \over \sim$}}

\def\ss{G0.17-0.41}
\def\ps{G359.55+0.16}
\def \ins{\hbox{\it Chandra}}
\def \xrism{\hbox{\it XRISM}}
\begin{document}

\lefttitle{Q. D. Wang}
\righttitle{Magnetic Feedback}

\jnlPage{1}{7}
\jnlDoiYr{2026}
\doival{10.1017/xxxxx}

\aopheadtitle{Proceedings IAU Symposium}
\editors{M. Zaja\v{c}ek,  T. Je\v{r}\'{a}bkov\'{a}, V. Karas, R. Schödel \&  P. Sukov\'{a}, eds.}

\title{X-ray thread/Nonthermal Radio Filament associations: Evidence for Interstellar Magnetic Reconnection}

\author{Q. Daniel Wang}
\affiliation{University of Massachusetts, USA}

\begin{abstract}
Nonthermal radio filaments (NTFs), first discovered at 20-centimeter wavelength more than four decades ago, are among the most enigmatic structures at the Galactic Center. They still defy a clear explanation. These striking narrow features trace intense magnetic fields and often stand in bold contrast to the Galactic plane. Recent discoveries have revealed surprising associations: some NTFs align well with X-ray threads that seem to exhibit Fe He-$\alpha$ emission. Here, I present preliminary results from an ongoing, collaborative, multi-wavelength study aimed at understanding the origins of these filaments, focusing on testing the magnetic reconnection scenario of these associations and shedding new light on the high-energy processes and magnetic phenomena operating under extreme conditions at the heart of our Galaxy.
\end{abstract}


\maketitle

\section{Introduction}

Radio and X-ray observations of the Galactic Center (GC) have uncovered a variety of distinctive energetic structures \citep[e.g., ][]{Heywood2019,Wang2021}, among them many thin radio filaments \citep{Yusef-Zadeh2022}. These features appear to mainly ``radiate" outward from the most active region of the central molecular zone (CMZ). Several have been found to exhibit strong polarization \citep[e.g., ][]{Yusef-Zadeh1984,Larosa2001}, indicating that their emission is largely synchrotron in nature; therefore, they are often called nonthermal filaments (NTFs). 

The physical processes responsible for the formation of NTFs remain poorly understood \citep[e.g., ][]{Heywood2019,Yusef-Zadeh2019,Sofue2020,Wang2021}.
Some of them appear to arrange into clusters that resemble harps \citep[e.g., G0.13-0.11; ][]{Wang2002,Heywood2022}. Such configurations can be naturally explained if magnetic flux tubes are illuminated by moving energetic sources, such as pulsars, that drive powerful winds of relativistic electrons and positrons. Indeed, G0.13-0.11, a curved X-ray thread that wraps around a "harp knee” of NTFs, hosts a point-like X-ray source, most likely representing a young pulsar \citep[e.g.,][]{Wang2002,Zhang2020}. G0.13-0.11 is also associated with TeV emission, which can be generated when pulsar-wind particles boost intense infrared radiation in the GC via inverse Compton scattering. In this framework, the magnetic flux tubes traced by the NTF ensemble may arise from the magneto-hydrodynamic pumping instability \citep{Bellan2003}, triggered by electric currents produced as pulsar wind particles stream in a charge-dependent manner along pre-existing magnetic field lines \citep{Bandiera2008,Olmi2019,Wang2021jet}. This instability compresses the plasma into self-contained magnetic flux tubes with a longitudinally uniform cross section. 

\begin{figure*}[htb!]
\centerline{
\includegraphics[width=0.95\textwidth]{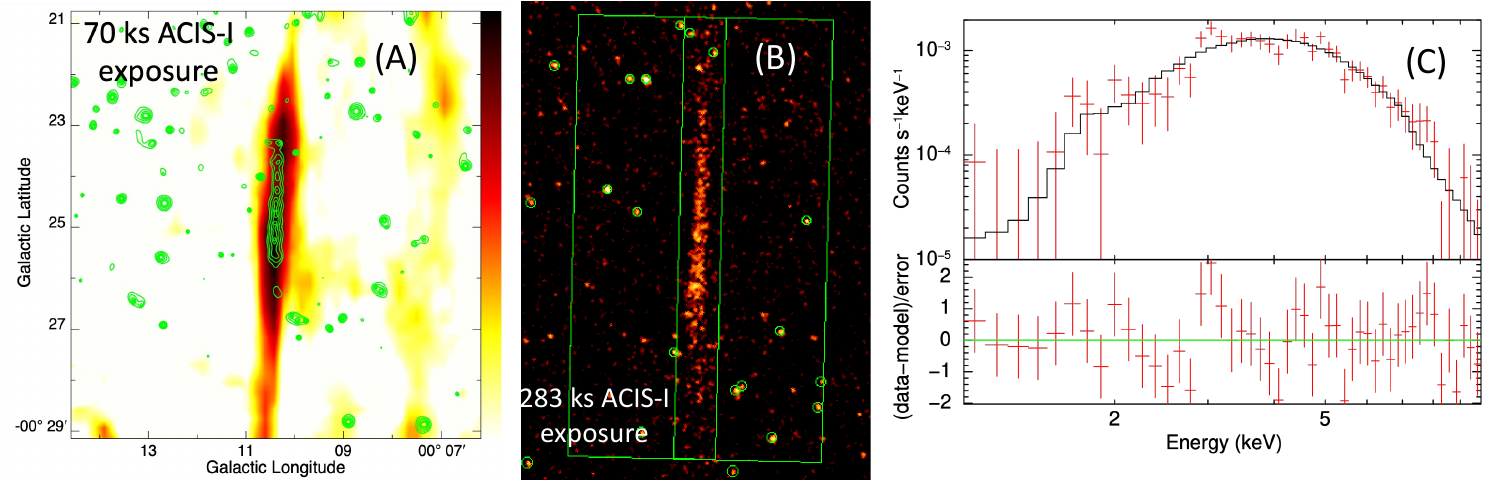}
}
\caption{X-ray spatial and spectral representation of \ss: {\bf (A)} Comparison with the MeerKAT 1.3-GHz image, overlaid with Chandra/ACIS-I 2.5–6 keV intensity contours  (adapted from \cite{Wang2021}), displayed in Galactic coordinates.
{\bf (B)} A preliminary merged Chandra image incorporating a new deep ACIS-I exposure; the boxes outline the extraction regions for the on-feature spectrum (central narrow box) and the off-feature background spectra. {\bf (C)} APEC plasma model fit to the net ACIS-I spectrum (Table~\ref{t:spec}).}
\label{f:f1}
\end{figure*}

\begin{figure*}[htb!]
\centerline{
\includegraphics[width=0.95\textwidth]{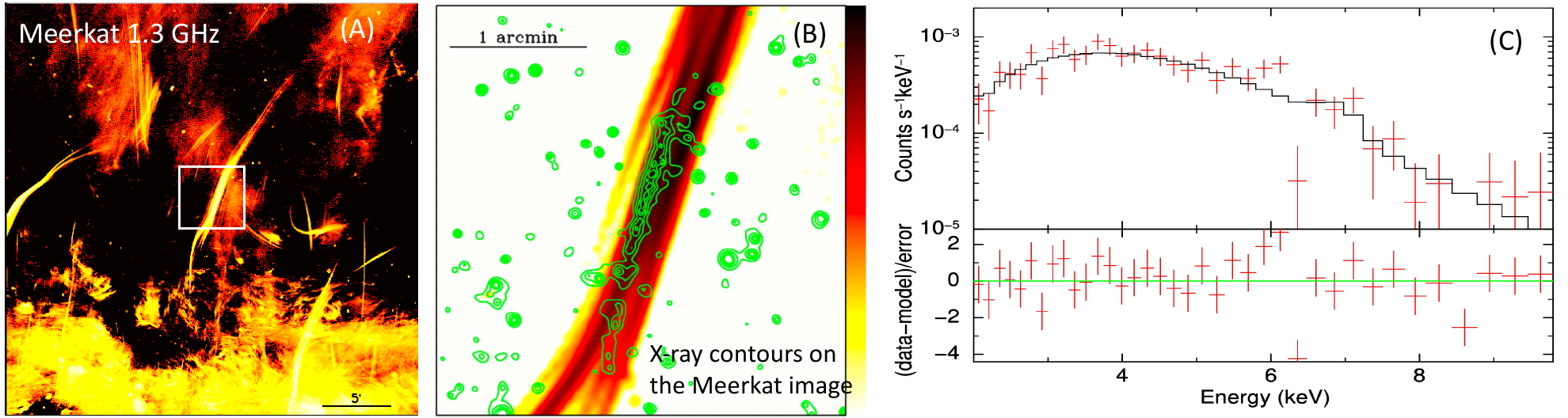}
}
\caption{X-ray spatial and spectral characterization of \ps: {\bf (A)} An overview of the \ps\ and its large-scale vicinity in the MeerKAT 1.3~GHz map \citep{Heywood2019}, displayed in Galactic coordinates. {\bf (B)} Zoomed-in view of \ps\ in the MeerKAT map overlaid with 4–9~keV ACIS-I intensity contours. {\bf (C)} APEC thermal plasma model fit to the net ACIS-I spectrum of the thread (see Table~\ref{t:spec}).
}
\label{f:f2}
\end{figure*}

However, pulsar-based interpretation encounters challenges in accounting for some other X-ray thread/NTF associations, particularly \ss\ \citep[Fig.~\ref{f:f1}; ][]{Wang2021} and \ps\ \citep[Fig.~\ref{f:f2}; ][]{Lu2003,Johnson2009}. This difficulty arises from their distinct morphological and/or spectral characteristics, as well as the absence of any obvious energetic point-like counterparts in radio or X-ray \citep{Wang2021}. In particular, \ps\ apparently shows a prominent FeXXV-He$\alpha$ 6.7-keV emission line with an equivalent width of $\sim 0.9^{+0.4}_{-0.4}$ keV (90\% confidence interval) in a 141~ks \suzaku\ observation, indicating that the X-ray emission is predominantly thermal in origin \citep{Yamauchi2014}, which is not expected in the pulsar-based interpretation.

Magnetic reconnection (MR) has been proposed to power \ss\ and \ps\ \citep{Lu2003,Yusef-Zadeh2021,Wang2021}. In this framework, these two X-ray thread/NTF systems -- and potentially additional NTFs -- can be interpreted in a unified way. MR occurs at the boundary between ionized or partially ionized gas regions in which the magnetic field lines are directed oppositely \citep[e.g.,][]{Lazarian2020}. Under such conditions, a fractal tearing instability of the field lines can occur, giving rise to plasmoids \cite[magnetic islands; e.g.,][]{Furth1963}, in which a large fraction of the magnetic energy is converted into heat through ohmic dissipation. The bulk of the X-ray emission may be produced by individual plasmoids. The electric field induced by the varying magnetic field, together with collisions among such plasmoids, can further accelerate particles, thereby generating cosmic rays \citep[e.g.,][]{Bicknell2001}. The heated plasma is ultimately expelled from the MR sites by magnetic tension associated with the strong curvature of the reconnected field lines. Therefore, interstellar MR could play a major role in heating plasma, accelerating cosmic rays, driving turbulence, and regulating global ISM structures. The X-ray thread/NTF pairings of \ss\ and \ps\ -- and in particular the likely detection of the 6.7 keV emission line in the latter -- arguably constitute the strongest observational evidence so far for MR operating in the interstellar medium.

\section{Recent X-ray observations and preliminary results}
We have obtained deep on-axis \ins\ observations of \ss\ and \ps\ (each with a $\sim$210 ks exposure) to better characterize their X-ray properties (Figs.~\ref{f:f1}B-C and \ref{f:f2}). Compared to the datasets used in previous analyses \citep[e.g., Fig.~\ref{f:f1}A; ][]{Wang2021,Lu2003}, the counting statistics have improved substantially, enabling a more detailed investigation of the spatial and likely temporal characteristics of the structures. In this work, we focus on preliminary spectral analysis, although the data's spectral resolution is intrinsically low, partly due to the increasing charge-transfer inefficiency (CTI) in the CCD detector. CTI is commonly expressed as the fractional charge loss per pixel transfer and produces two main observable effects: a deterioration in energy resolution (FWHM) and a reduction in quantum efficiency (as some events are shifted into excluded grades). For the Fe 6.7-keV emission line, for instance, CTI causes both an apparent decrease in the line energy (a gain loss) and a broadening of the line profile, with these effects becoming progressively stronger over the course of the mission. At the near on-axis position of our new ACIS-I observations, a representative FWHM for this line is $\sim 200$\,eV in the new observations. 

We perform X-ray spectral analysis of \ps\ and \ss\ using FTOOLS software packages, particularly XSPEC for spectral fitting. This analysis is based on the spectra grouped with the task ftgrouppha, adopting the optimal binning prescription of \citet{Kaastra2016} with an additional constraint on the minimum signal-to-noise ratio (i.e., grouptype=optsmin and groupscale=3). Our tests indicate that the fitting results are not sensitive to this particular choice of grouping strategy. We fit the spectra using the thermal plasma model assuming collisional ionization equilibrium (CIE). The fit results are summarized in Table~\ref{t:spec}.
\begin{table}[!htb]
 \centering
\leavevmode
\caption{APEC spectral characterization of \ss\ and \ps}\label{t:spec}
\begin{tabular}{lcc}
 \midrule 
Parameter & \ss\ & \ps\ \\
 \midrule 
$N_H (10^{22} {\rm~cm^{-2}})$ & 5.9 (4.7-7.8) & 5.2 (4.0-6.5)\\ 
$kT$(keV)  & 17 (6.0-)  & 9.7 (5.8-23) \\
$Z$ (solar) & 0.15 (0-3.3) & 0.58 (fixed) \\
Flux$^{a,b}$  &1.2 & 0.77\\
Log($L_x$)$^b$ &  33.2 (33.1-33.3) & 32.9 (32.8 - 33.0)\\
$\chi^2/d.o.f.$ & 43.3/46 & 52.1/34 \\
  \midrule
\end{tabular}

$^a$ The absorbed flux is in units of $10^{-13} {\rm~ergs~s^{-1}~cm^{-2}}$.
$^b$ Both flux and luminosity are estimated in the 2-8~keV band; the latter assumes a distance of 8.2 kpc.
\end{table}

It is instructive to compare the results of the spectral fitting obtained with \chandra\ and \suzaku\ for \ps. Due to the limited spectral quality of the \chandra\ data, the metal abundance $Z$ cannot be meaningfully constrained. We therefore fix $Z$ to 0.58, the best-fit value derived from the \suzaku\ spectrum, to allow a direct comparison of the remaining parameters between the two results. The resulting null-hypothesis probability of 0.02 indicates a rather unsatisfactory fit. As illustrated in the residual panel of Fig.~\ref{f:f2}C, the most prominent mismatch between the model and the data occurs near $\sim 6$\,keV, which is physically difficult to explain. Although the best-fit $N_H$ is consistent with the value of \suzaku\ \citep[$N_H = 6.1^{+2.5}_{-1.3}$;][]{Yamauchi2014}, the fitted temperature $kT$ is considerably higher than that obtained from \suzaku\ ($4.1^{+2.7}_{-1.8}$). However, the associated uncertainties are large enough that no robust statement can be made. 

The most striking disparity between the two results lies in the inferred fluxes: 
$0.77 \times 10^{-13} {\rm~erg~s^{-1}~cm^{-2}}$ from \chandra\ versus $1.8\times 10^{-13} {\rm~erg~s^{-1}~cm^{-2}}$ from \suzaku. This factor-of-2.3 discrepancy, which is largely insensitive to the specific spectral model adopted, can be attributed at least in part to differences in the spectral extraction regions. To test this hypothesis, we extract a spectrum from an elliptical region with semi-minor and semi-major axes of 1$^\prime$ and 2$^\prime$, respectively -- matching the region used in the \suzaku\ analysis -- while choosing a background region located in approximately the same sky area as in the \suzaku\ study. When source removal is disabled to replicate the limited resolution of the \suzaku\ data, the resulting new \chandra\ spectrum is likewise well fitted by an APEC model, and the derived 2–8 keV flux agrees with the \suzaku\ measurement. This clearly demonstrates that contamination from nearby sources significantly affects the \suzaku\ spectrum of \ps, implying that its spectral-fit result must be interpreted with caution.

\begin{figure*}[htb!]
\centerline{
\includegraphics[width=0.95\textwidth]{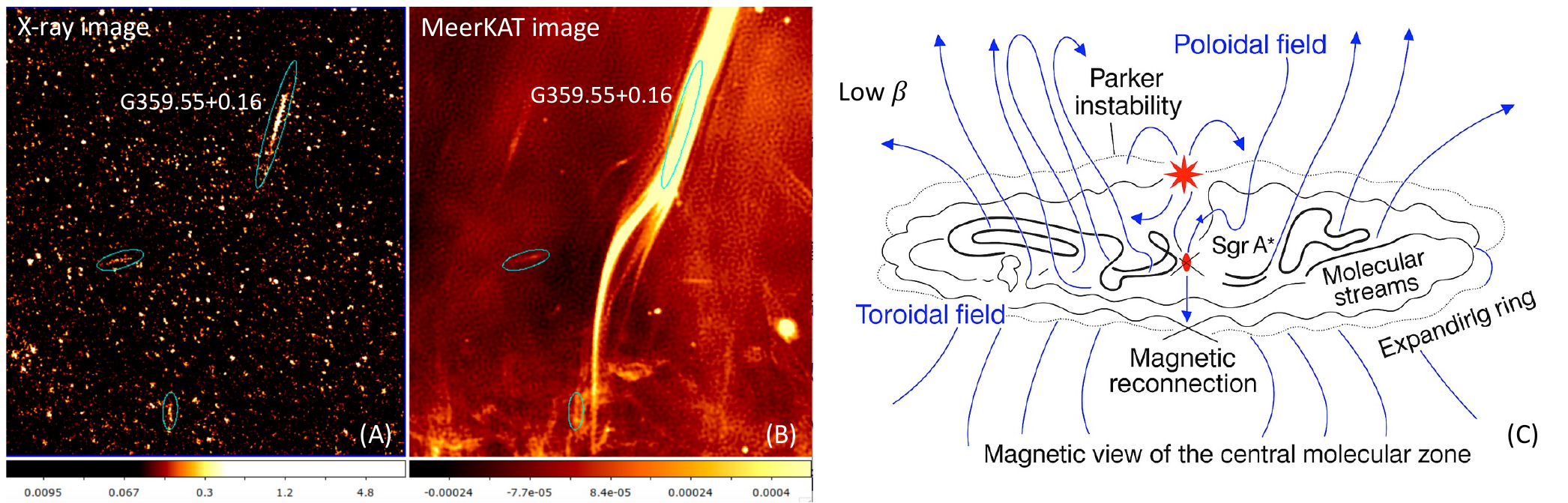}
}
\caption{Detection of two additional NTF/X-ray thread associations (as enclosed in two small green ellipses) near \ps, seen in the \chandra/ACIS-I image {\bf (A)} and the MeerKAT image {\bf (B)}. These two X-ray threads are too faint to permit a reliable spectral analysis with the existing data. {\bf (C)} Illustration of global magnetic structure in and around the CMZ and potential sites of MR events (marked in red). 
}
\label{f:f3}
\end{figure*}
 
 

We also identify two more X-ray threads located just a few arcminutes from \ps, each associated with an NTF counterpart(Fig.~\ref{f:f3}). These associations show no apparent point-like X-ray or radio sources. All thin X-ray threads identified so far have associated NTFs, but not vice versa.

\section{Discussion and Conclusion}

A definitive confirmation of the thermal nature of the X-ray threads, though still pending, would strongly affect our understanding of the nature of X-ray thread/NTF associations and the heating of the interstellar medium (ISM). Hot thermal plasma production by MR in the ISM is theoretically predicted by analogy with solar coronal heating \citep[e.g, ][]{Raymond1992,Tanuma2003,Florido-Llinas2020} and invoked to explain diffuse X-ray emission in nearby galaxies \citep[e.g.,][]{Wezgowiec2020} and in the Milky Way \citep[e.g.,][]{Tanuma2003}. MHD simulations \citep{Tanuma2003, Hanasz2002} show that MR can occur around expanding shells (e.g., from Parker or magnetic buoyancy instabilities) or in the wakes of molecular clouds interacting with a global wind \citep[e.g., Fig.~\ref{f:f3}C; ][]{Banda-Barragan2016}. Thus, conclusively establishing the thermal origin of the X-ray emission from \ps\ and \ss\ would be a critical test of the MR scenario.

To achieve this, we require deep observations of \ss\ and \ps\ with \xmm\ and \xrism. \xmm/\textit{EPIC} will deliver the counting statistics and broad energy coverage needed to map the X-ray spectral distribution with moderate spatial and spectral resolution. XRISM/\textit{Resolve} will offer the exceptional spectral resolution necessary to robustly detect and characterize the line emission (Fe\,\textsc{XXV}, Fe\,\textsc{XXVI}, and 6.4\,keV fluorescence). These data will allow us to spectroscopically disentangle the relative contributions of the X-ray threads (likely diffuse thermal plasma) and the nearby excess of cataclysmic variables and/or active binaries. 
Combined with the existing \chandra\ observations, which already place tight constraints on the geometry of the threads and on the spatial distributions of nearby X-ray sources, this synergy will enable us to definitively determine whether the threads are thermal or nonthermal and investigate their physical and potentially chemical properties. If the threads are confirmed to be thermal, a joint analysis of all datasets will allow us to test the CIE assumption and to measure both the abundance of Fe and the temperature ($kT$) of the threads. This directly probes the solar/supersolar Fe abundances expected in the GC ISM and, in the \ps\ case, avoids the artificially low abundances that can result from relying solely on \suzaku\ data, given its limited resolution. Using the \textit{Chandra-constrained} volume of each feature, we can further convert its emission measure into a density $n$. With $(n, kT)$ tightly constrained, we can then estimate the magnetic field associated with MR heating, $B \simeq \sqrt{kT({\rm keV})\,n({\rm cm^{-3}})/10}$ mG \citep{Tanuma2003}, and derive the plasma mass, thermal pressure, and energy. Comparing the plasma cooling time with a characteristic advective timescale (e.g., one based on an Alfvénic speed) quantifies the energy injection rate due to MR. Such diagnostics have not previously been obtained for interstellar MR.

The detected X-ray thread/NTF associations probably represent just the tip of the MR iceberg in the GC \citep{Wang2021}. Some of the other NTFs could also be due to MR. Their X-ray emission may simply be too weak and/or too diffuse to be easily detected individually \citep{Tanuma2003}. The vicinities of \ss\ and \ps, outside the crowded CMZ, are ideal fields for this study, minimizing potential confusion with other diffuse and point-like X-ray sources. With higher sensitivities and larger fields of view, future deep \xmm\ EPIC imaging data will allow us to characterize the spectral properties of weak X-ray threads, as well as strong ones like \ps\ and \ss, individually and statistically. This will have immediate implications for understanding diffuse X-ray emission in the GC and will help determine how widespread MR is and how it may regulate the global ISM in the GC.

The author of this work expresses gratitude to the collaborators involved in the research presented here, with particular thanks to Mark Morris, Gabriele Ponti, and Shuo Zhang, who participate in ongoing projects that are partially funded by NASA Chandra Grants GO3-24120X and GO4-25091X. 



\bibliographystyle{aasjournal}
{\small
\bibliography{export-bibtex}
}

\end{document}